\documentstyle[pre,aps,multicol,epsf]{revtex}
\begin{document}
\draft
\widetext

\title {Wave Scattering through Classically Chaotic Cavities in the Presence of 
Absorption:
An Information-Theoretic Model}

\author{Eugene Kogan${}^{1}$, Pier A. Mello${}^{1,2}$ and 
He Liqun${}^{1,3}$}

\address{${}^1$Minerva Center and Jack and Pearl Resnick Institute of Advanced 
Technology,\\
Department of Physics, Bar-Ilan University, Ramat-Gan 52900, Israel}

\address{${}^2$Instituto de F\'{\i}sica, Universidad Nacional Aut\'onoma de 
M\'exico, 
M\'exico, D F}

\address{${}^3$Department of Thermal Science and Energy Engineering,\\ 
University of Science and Technology of China, Hefei, P.R.China}
 
\date{\today }

\maketitle
\widetext

\begin{abstract}
We propose an information-theoretic model for the transport of waves through
a chaotic cavity in the presence of absorption.
The entropy of the $S$-matrix statistical distribution is maximized, with  
the constraint $\left\langle {\rm Tr}SS^{\dagger}\right\rangle =\alpha n$:
$n$ is the dimensionality of $S$, and
$0\leq \alpha \leq 1$, $\alpha =0(1)$ meaning complete (no) absorption. 
For strong absorption our result agrees with a number of analytical 
calculations
already given in the literature.
In that limit, the distribution of the individual (angular) transmission and 
reflection coefficients becomes exponential --Rayleigh statistics-- even for 
$n=1$. 
For $n\gg 1$ Rayleigh statistics is attained even with no absorption; here
we extend the study to $\alpha <1$. 
The  model is compared with random-matrix-theory 
numerical simulations: it describes the problem very well for strong 
absorption, but fails for moderate and weak absorptions. 
Thus, in the latter regime, some important physical constraint is missing in the 
construction of the model.
\end{abstract}
\pacs{PACS numbers: 05.45.+b,42.25.Bs,41.20.Jb}

\begin{multicols}{2}
\narrowtext

Systems involving multiple elastic scattering of any kind of waves (sound,
microwaves or light) --whose interference leads to strong 
fluctuations in the transmitted intensity-- can be described in very much the 
same way 
as electron systems. However, there exists an important difference: the 
interference
pattern for classical waves can be affected as a result of loss or 
{\it absorption}, which is absent in electron systems.
The issue is extremely important from an experimental point of view, because
absorption is always present and is often very strong.  For diffusive
transport the problem was intensively studied both experimentally \cite{azi}
and theoretically \cite{pnini,kogan,brouwer}. The issue has also attracted
attention in connection with the 
phase-coherent reflection of light by a disordered medium which amplifies 
radiation \cite{beenakker,beenakker2} and the study of the relation between 
absorption and dephasing \cite{brouwer,brouwer2}.

The analytical evaluation of the reflection-matrix statistical distribution for 
a 
semi-infinite disordered waveguide
was performed, for arbitrary absorption, in Ref. \cite{beenakker},
and that of the $S$-matrix distribution for a chaotic cavity 
with absorption and one propagating mode in each of two waveguides, in Ref. 
\cite{brouwer2}.

In the present paper we take up again the problem of the propagation of scalar 
waves 
traveling through a cavity --whose classical dynamics would be chaotic-- 
in the presence of absorption, connected to the outside through a number of 
waveguides 
supporting an {\it arbitrary} number of propagating modes. Motivated by the success of 
an information-theoretic approach \cite{mello-baranger99} to the study of 
chaotic scattering 
through cavities, we propose below 
an extension of such models to study the effect of absorption. 
Since these models are based on the idea of doing statistics directly on the $S$ 
matrix 
of the system --on the basis of the information which is physically relevant 
for the problem in question-- we believe that the
present approach complements the analytical derivations mentioned above. 
We show that the two approaches agree in the limit of strong absorption,
while for moderate and weak absorption some  relevant information
is missed in our model.

The scattering of waves through a cavity can be described by an $S$ matrix that
relates incoming and outgoing amplitudes. The dimensionality $n$ of the
matrix is the total number of channels in all the waveguides. For two 
$N$-channel waveguides, $n=2N$ and the $S$ matrix has the structure 
\begin{equation}
S=\left[ 
\begin{array}{cc}
r & t^{\prime } \\ 
t & r^{\prime }
\end{array}
\right] ,  
\label{S}
\end{equation}
where $r,r^{\prime }$ and $t,t^{\prime }$ are the $N$-dimensional matrices
of reflection and transmission amplitudes with incidence from either
waveguide.

In quantum-mechanics, the universality classes for $S$
matrices were introduced by Dyson \cite{dyson,mehta}. In the absence of any
symmetries, the only restriction on $S$ is unitarity, $SS^{\dagger }=I$, due
to flux conservation (the {\it unitary} or $\beta =2$ case). In
the {\it orthogonal} case ($\beta =1$), $S$ is symmetric because of
either time-reversal invariance (TRI) and integral spin, or TRI,
half-integral spin and rotational symmetry. In the {\it symplectic} case 
($\beta =4$), $S$ is self-dual because of TRI with half-integral spin and no
rotational symmetry. The intuitive idea of {\it equal-a-priori probabilities}
is expressed mathematically by the {\it invariant measure} on the matrix space 
under the symmetry operation for the class in question,
giving the {\it circular} orthogonal, unitary and symplectic ensembles (COE, 
CUE, CSE).

Consider the orthogonal case. The potential
appearing in the Schr\"{o}dinger equation is real, with a strength $u_{0}$, 
say. The resulting $S$ matrix is unitary, $SS^{\dagger }=I$, and
symmetric, $S=S^{T}$. Suppose we analytically continue $u_{0}$ to complex 
values: 
$u_{0}=u_{0}^{\prime}-iu_{0}^{\prime \prime }$, the sign of the imaginary part 
ensuring
absorption. Because of loss of flux, the resulting $S$ is now 
{\it subunitary}, meaning that the eigenvalues of the Hermitian matrix 
$h=SS^{\dagger }$ lie in the interval between zero and one. However, the 
{\it symmetry} property, $S=S^{T}$, is not altered. 
We still speak of the orthogonal ($\beta =1$) case.
For the unitary one, a similar analytic continuation gives a subunitary and in
general non-symmetric $S$ matrix. In the scattering problem of scalar classical 
waves, the orthogonal case is the physically relevant one. However, we deal 
below 
with both $\beta =1$ and $2$, the unitary case being presented as a reference
problem, as it is often simpler to treat mathematically than the orthogonal
one.

Following Ref.\cite{hua} (pp. 63, 64), we introduce a {\it uniform weight in
the space of sub-unitary matrices} as
\begin{equation}
d\mu _{sub}^{(\beta )}(S)=C\theta (I-SS^{\dagger })\prod_{a,b}dx_{ab}dy_{ab},
\label{dmu}
\end{equation}
where $S_{ab}=x_{ab}+iy_{ab}$; 
$\prod_{a,b}$ is over all elements in the unitary case, but, in
the orthogonal one, only over $a \leq b$. In this equation, the step function 
$\theta (H)$ (for a Hermitian $H$) is nonzero for $H>0$ (i.e.  
for $H$ positive definite, so that all its eigenvalues are positive definite) 
and thus selects {\it sub-unitary} matrices.

A complex $n\times n$ sub-unitary matrix can be written in the {\it polar
representation} (Refs. \cite{hua}, pp. 63,64 and \cite{brouwer2}) as 
\begin{equation}
S=UDV.  
\label{S polar}
\end{equation}
The unitary matrices $U$ and $V$ are arbitrary in the unitary case, while 
$V=U^T$ in the orthogonal one. The matrix $D$ is diagonal, with the structure 
$D_{ab}=$ $\rho _a\delta _{ab}$ ($a=1,\cdot \cdot \cdot ,n$), with 
$0\leq \rho _a\leq 1$.

The explicit expression of the above measure (\ref{dmu}) in terms of the
independent parameters of the polar representation (\ref{S polar}) is 
(Ref. \cite{hua}, pp. 63, 64) 
\begin{equation}
d\mu _{sub}^{(\beta )}(S)=\prod_{a<b}^{n}|\rho _{a}^{2}-\rho _{b}^{2}|^{\beta }
\prod_{c}\rho _{c}d\rho _{c}d\mu (U)d\mu (V),  
\label{dmu polar}
\end{equation}
where $d\mu (U)$, $d\mu (V)$ are the invariant measure for the unitary group
in $n$ dimensions; $d\mu (V)$ is absent in the orthogonal case $\beta =1$.

More general statistical distributions of sub-unitary matrices carrying more
information than the equal-a-priori probability (\ref{dmu}) can now be
constructed using $d\mu _{sub}(S)$ as a starting point and writing 
\begin{equation}
dP(S)=p(S)d\mu _{sub}(S).  
\label{dP}
\end{equation}
In what follows we propose an information-theoretic criterion to choose 
$p(S)$. The {\it information-theoretic entropy} ${\cal S}$
of the $S$-matrix distribution \cite{mello-pereyra-seligman},
${\cal S}\left[ p\left( S\right) \right] = -\int p(S)\ln p(S)d\mu _{sub}(S)$,
is {\it maximized subject to the constraint of a given average strength of the 
absorption}.
Mathematically, this is expressed by the average departure from unitarity of
our $S$ matrices; we thus write the constraint as
\begin{equation}
\left\langle {\rm Tr}SS^{\dagger }\right\rangle =\alpha n,\;\;\;
0\leq \alpha \leq 1.  
\label{constraint}
\end{equation}
Thus $\alpha =0$ corresponds to complete absorption and $\alpha =1$ to lack
of absorption. 
We find 
\begin{equation}
dP^{(\beta )}(S)=Ce^{-\nu {\rm Tr}SS^{\dagger }}d\mu _{sub}^{\beta }(S),  
\label{dP 1}
\end{equation}
where the constant $C$ and the Lagrange multiplier $\nu $ ensure normalization
and the fulfillment of the constraint (\ref{constraint}).
The limit of no absorption, $\alpha =1$, is attained when the Lagrange 
multiplier 
$\nu \rightarrow -\infty $ and the distribution concentrates on the unitarity 
sphere. 
The limit of complete absorption, $\alpha \rightarrow 0$, is attained when 
$\nu \rightarrow +\infty $: the distribution then becomes a $\delta $-function 
at the origin and there is no exit signal. 
The result of Eq. (\ref{dP 1}) --Laguerre ensemble for the variables 
$\rho _{a}^{2}$-- coincides, for strong absorption, with that obtained in 
Ref. \cite{beenakker} for the diffusive waveguide and in Ref. \cite{brouwer2}
for cavities with $N=1$.

The ``ansatz'' (\ref{dP 1}) entails a number of properties and restrictions.
>From (\ref{dmu polar}), (\ref{S polar}) and the properties of $d\mu (U)
$ \cite{mello90}, we see that, under the distribution (\ref{dP 1}), the
average $\left\langle S\right\rangle =0$. Therefore, applications of the
model (\ref{dP 1}) should be restricted to cases where {\it prompt processes}
are absent and so $\left\langle S\right\rangle =0$ 
\cite{mello-pereyra-seligman,baranger-mello96,mello-baranger99}.
We shall see at the end of the paper that for uniform volume absorption 
(in quantum mechanics, a constant potential $-iW$ throughout the cavity) the $S$ 
matrix
can be obtained from that without absorption, evaluating it at the complex
energy $E+iW$, i.e. $S(E+iW)$. 
The so-called analyticity-ergodicity requirements 
$\left\langle S_{a_{1}b_{1}}^{n_{1}}\cdot \cdot \cdot
S_{a_{p}b_{p}}^{n_{p}}\right\rangle 
=\left\langle S_{a_{1}b_{1}}\right\rangle ^{n_{1}}\cdot \cdot \cdot 
\left\langle S_{a_{p}b_{p}}\right\rangle ^{n_{p}}$
are discussed in Ref. \cite{mello-pereyra-seligman} for unitary matrices:
the same argument applies here as well, and the necessity to fulfill them
follows. In the present case, their fulfillment follows from (\ref{dP 1}), 
(\ref{dmu polar}), (\ref{S polar}) and the properties of $d\mu (U)$ 
\cite{mello90}.

As a brief excursion into the situation when 
$\left\langle S \right\rangle \neq 0$, we note that,
as $W$ increases, $S(E+iW)\rightarrow \left\langle S\right\rangle $, 
which, in turn, is the so called
``optical $S$ matrix'', that signals the presence of prompt or direct
processes \cite{mello-pereyra-seligman,baranger-mello96,mello-baranger99}. 
As a consequence, 
$S$, for large absorption, does not tend to zero, but to 
$\left\langle S \right\rangle $.  With no absorption,
the splitting $S=\left\langle S\right\rangle +S^{fl}$ describes the problem
in terms of two responses, associated with two distinct time scales: the
prompt and the equilibrated one. It is then expected that strong absorption
will affect the former response, that corresponds to short trajectories,
much less then the latter one, that arises from long chaotic trajectories.

We analyze some of the consequences of the ansatz (\ref{dP 1}) and, at
the end, compare them with the results of random-matrix-theory (RMT)
numerical simulations.

\paragraph{The $n=1$ case.}
This case, which describes a cavity with one waveguide supporting only one open 
channel
($S$ is thus the reflection amplitude back to the only channel we have),
is, within our model (\ref{dP 1}), independent of the universality class $\beta 
$. 
Eq. (\ref{S polar}) for $S$ in the polar representation reduces to
$S=\rho \exp {i\theta }$; $\rho ^{2}$ represents the reflection coefficient $R$. 
The uniform weight (\ref{dmu polar}) and the distribution (\ref{dP 1}) reduce
to
\begin{equation}
d\mu _{sub}(S)=\rho d\rho d\theta ,  \; \; \; dP(S)=Ce^{-\nu \rho ^{2}}\rho 
d\rho d\theta .
\label{dP n1}
\end{equation}
The $R$-probability density is
\begin{equation}
w(R)=De^{-\nu R},\qquad 0\leq R\leq 1, 
\label{w(R)}
\end{equation}
$D$ and $\nu $ being given by
\begin{equation}
D=\frac{\nu }{1-e^{-\nu }}, \; \; \;
\left\langle R\right\rangle =\frac{1}{\nu }-\frac{1}{e^{\nu }-1}=\alpha .
\label{D}
\end{equation}
For weak absorption, $\alpha \approx 1$, $\nu \rightarrow -\infty $
and the distribution (\ref{w(R)}) becomes strongly peaked around 
$R=1$, i.e. the unitarity circle, reducing to the one-sided delta function 
$\delta (1-R)$ as $\alpha \rightarrow 1$. In the other extreme of
{\it strong absorption}, $\nu \rightarrow +\infty $, $\alpha \approx 1/\nu $, 
$D\approx 1/\alpha $ and
\begin{equation}
w(R)\approx \left\langle R \right\rangle ^{-1} 
e^{-R/\left\langle R \right\rangle},
\label{w(R)strong abs}
\end{equation}
{\it Rayleigh's distribution}, with the average 
$\left\langle R \right\rangle = \alpha $.

\paragraph{The orthogonal case for arbitrary $n$.}
Using the results of Ref. \cite{mello90} we find, for the average of an 
individual (angular) transmission or reflection coefficient
\begin{equation}
\left\langle T_{ab}\right\rangle ^{(1)}
=\left\langle R_{a\neq b}\right\rangle ^{(1)} 
= \alpha /(n+1)
= (1/2) \left\langle R_{aa} \right\rangle ^{(1)}
\label{av(Tab)orth}
\end{equation}
We see the occurrence of the familiar {\it backward enhancement factor 2.}
For the second moments we find
\begin{eqnarray}
\left\langle T_{ab}^{2}\right\rangle ^{(1)}
=\left\langle R_{a\neq b}^{2}\right\rangle ^{(1)} 
=\sum_{\alpha =1}^{n}
\left\langle |U_{1\alpha }|^{4} |U_{2\alpha}|^{4} \right\rangle _{0}
\left\langle \rho _{\alpha }^{4}\right\rangle ^{(1)} \nonumber \\
+ 2\sum_{\alpha \neq \gamma =1}^{n}
\left\langle |U_{1\alpha }|^{2} |U_{2\alpha }|^{2}
|U_{1\gamma }|^{2} |U_{2\gamma }|^{2}
\right\rangle _{0}
\left\langle \rho _{\alpha }^{2} \rho _{\gamma }^{2}\right\rangle ^{(1)}.
\label{<Tab2>orth}
\end{eqnarray}
The indices 1 and 2 indicate any pair of different channels;
$\left\langle \cdot \cdot \cdot \right\rangle _{0}$ stands for an average with 
respect to the invariant measure of the unitary group \cite{mello90}.
For $\left\langle R_{aa}^{2}\right\rangle ^{(1)}$ one sets the two
indices $1$ and $2$ equal. 

For a two-waveguide problem with one channel in each waveguide 
the $S$ matrix is two dimensional ($n=2N=2$).
Eq. (\ref{av(Tab)orth}) gives 
$\left\langle T \right\rangle ^{(1)} = \alpha /3$, 
$\left\langle R \right\rangle ^{(1)}= 2\alpha /3$. 
Restricting ourselves to the limit of {\it strong absorption}, $\alpha \ll 1$, 
the
Lagrange multiplier $\nu =3/2\alpha $ and we obtain 
\begin{equation}
\left\langle T^{2} \right\rangle ^{(1)}
\approx 2\left[ \left\langle T\right\rangle^{(1)}\right] ^{2}.  
\label{Rayl n2 orth}
\end{equation}
Although we have not calculated the statistical distribution of $T$, result 
(\ref{Rayl n2 orth}) is consistent with the 
Rayleigh distribution for the transmission coefficient $T$.

For a large number of channels, $n=2N\gg 1$, Eqs. (\ref{<Tab2>orth}) and 
(\ref{av(Tab)orth}) give \cite{mello90}
\begin{equation}
\left\langle T_{ab}^{2}\right\rangle ^{(1)}
=\left\langle R_{a\neq b}^{2}\right\rangle ^{(1)}
\approx 2\frac{\alpha ^{2}}{n^{2}}
\approx 2\left[\left\langle T_{ab} \right\rangle ^{(1)}\right] ^{2}.
\label{RaylTab n>>1orth}
\end{equation}
and similarly, $\left\langle R_{aa}^{2} \right\rangle ^{(1)}
\approx 2\left[ \left\langle R_{aa} \right\rangle ^{(1)}\right] ^{2}$,
the relation between first and second moments for an exponential distribution 
with the average value (\ref{av(Tab)orth}), which becomes smaller
as the absorption increases.

For no absorption one reaches, in the limit $n\gg 1$, a Rayleigh
distribution for $R_{aa}$. 
Ref. \cite{pereyra-mello83} shows that, for the COE
($\approx $ being an approximation for large $n$)
\[
w^{(1)}(R_{aa})
=C \left( 1-R_{aa}\right) ^{\frac{n-3}{2}}
\approx \left\langle R_{aa} \right\rangle ^{-1}
e^{-R_{aa}/\left\langle R_{aa} \right\rangle}.
\]

\paragraph{The unitary case for arbitrary $n$.}
In the unitary case, the statistical properties of a transmission
coefficient is identical to that of a diagonal or off-diagonal reflection
coefficient. The following equations are thus written for 
$T_{ab}$. We find, for its average,
$\left\langle T_{ab}\right\rangle ^{(2)} = \alpha /n$;
we have used the result \cite{mello90} 
$\left\langle \left| U_{1\alpha }\right| ^{2}\right\rangle _{0} = 1/n$.
The difference in expectation values between the two symmetry classes is thus
\begin{equation}
\left\langle T_{ab}\right\rangle ^{(1)}-\left\langle T_{ab}\right\rangle ^{(2)}
=-\alpha /[n(n+1)].  
\label{WLC}
\end{equation}

For the second moment of $T_{ab}$ we have 
\begin{eqnarray}
\left\langle T_{ab}^{2}\right\rangle ^{(2)} = 2\sum_{\alpha \gamma } 
\left\langle |U_{1\alpha }|^{2}|U_{1\gamma }|^{2}\right\rangle _{0}
\left\langle |V_{\alpha 2}|^{2}|V_{\gamma 2}|^{2}\right\rangle _{0}
\left\langle \rho _{\alpha }^{2}\rho _{\gamma }^{2}\right\rangle  \nonumber \\
=[2/(n+1)^{2}]
\left\{ [(n-1)/n] \left\langle \rho _{1}^{2} \rho _{2}^{2}\right\rangle 
^{(2)}
+(2/n)\left\langle \rho _{1}^{4}\right\rangle ^{(2)}\right\} .  
\label{<Tab2>unit}
\end{eqnarray}
We have used the result \cite{mello90} 
$\left\langle |U_{1\alpha }|^{2}|U_{1\gamma }|^{2}\right\rangle _{0}
=\left(1+\delta _{\alpha \gamma} \right)/\left[n(n+1)\right]$. 

For a two-waveguide problem with one channel in each waveguide ($n=2N=2$)
we have $\left\langle T\right\rangle ^{(2)}=\alpha /2$.
In the limit of {\it strong absorption}, $\alpha \ll 1$, the
Lagrange multiplier $\nu =2/\alpha $ and we obtain a relation
like (\ref{Rayl n2 orth}).
For a large number of channels $n=2N\gg 1$, Eq. (\ref{<Tab2>unit}) gives
a similar relation, now for $T_{ab}$, which is again consistent with 
Rayleigh's distribution.

For CUE (i.e. no absorption) one reaches a Rayleigh distribution for 
$n\gg 1$. Ref. \cite{pereyra-mello83} finds for the
the distribution of a single transmission coefficient as 
\begin{equation}
w(T_{ab}) = C \left( 1-T_{ab}\right) ^{n-2}
\approx \left\langle T_{ab} \right\rangle ^{-1} 
e^{-T_{ab}/\left\langle T_{ab} \right\rangle}.
\label{w(Tab)unit largeN no abs}
\end{equation}

\paragraph{Comparison with numerical simulations}

Some of our predictions are compared below with RMT numerical simulations 
for $\beta =1$. The $S$ matrices are constructed as 
$S(E)=-\left[I_{n}-iK(E)\right] ^{-1} \left[I_{n}+iK(E)\right]$,
with
$K_{ab}(E)
=\sum_{\lambda }\gamma _{\lambda a}\gamma _{\lambda b}/
\left[E_{\lambda }-E\right]$
and $(I_{n})_{ab}=\delta _{ab}$ ($a,b=1,\cdot \cdot \cdot ,n$).
The $E_{\lambda }$'s are generated from an ``unfolded'' zero-centered GOE 
\cite{brody et al} with average spacing $\Delta$.
The $\gamma _{\lambda a}$'s are 
statistically independent, real, zero-centered Gaussian random variables.
At $E=0$, $\left\langle S_{ab} \right\rangle 
=-\left[1+\pi \left\langle \gamma _{\lambda a}^{2}\right\rangle /\Delta \right] ^{-1}
\left[1-\pi \left\langle \gamma _{\lambda a}^{2}\right\rangle /\Delta \right]
\delta _{ab}$, and we require $\left\langle S \right\rangle = 0$. 
In the quantum case, addition of a constant imaginary potential $-iW$ inside
the cavity makes the $E_{\lambda }$'s complex and equal to $E_{\lambda }-iW$
(see also Ref. \cite{brouwer2}). 
This is equivalent to evaluating the above expressions at the complex
energy $E+iW$, which makes $S(E+iW)$ subunitary. 
Although Eq. (\ref{dP 1}) gives the probability distribution for the full $S$ matrix
and arbitrary $n$,
we only analyze below individual (angular) reflection and transmission coefficients,
for $n=1,2$.
\begin{figure}[tbh]
\epsfxsize=4cm
\centerline{\epsffile{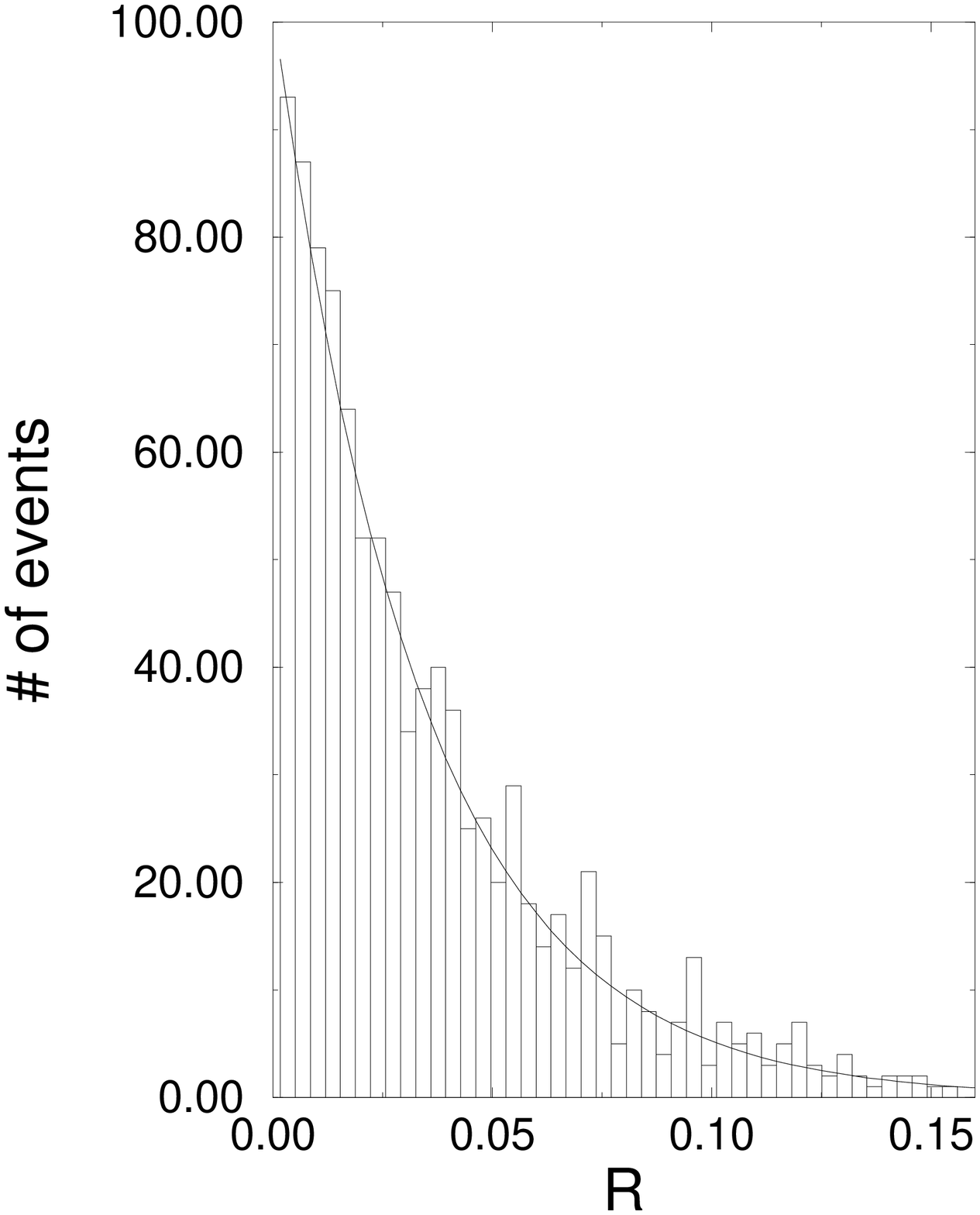} \hskip .3 cm \epsfxsize=4cm \epsffile{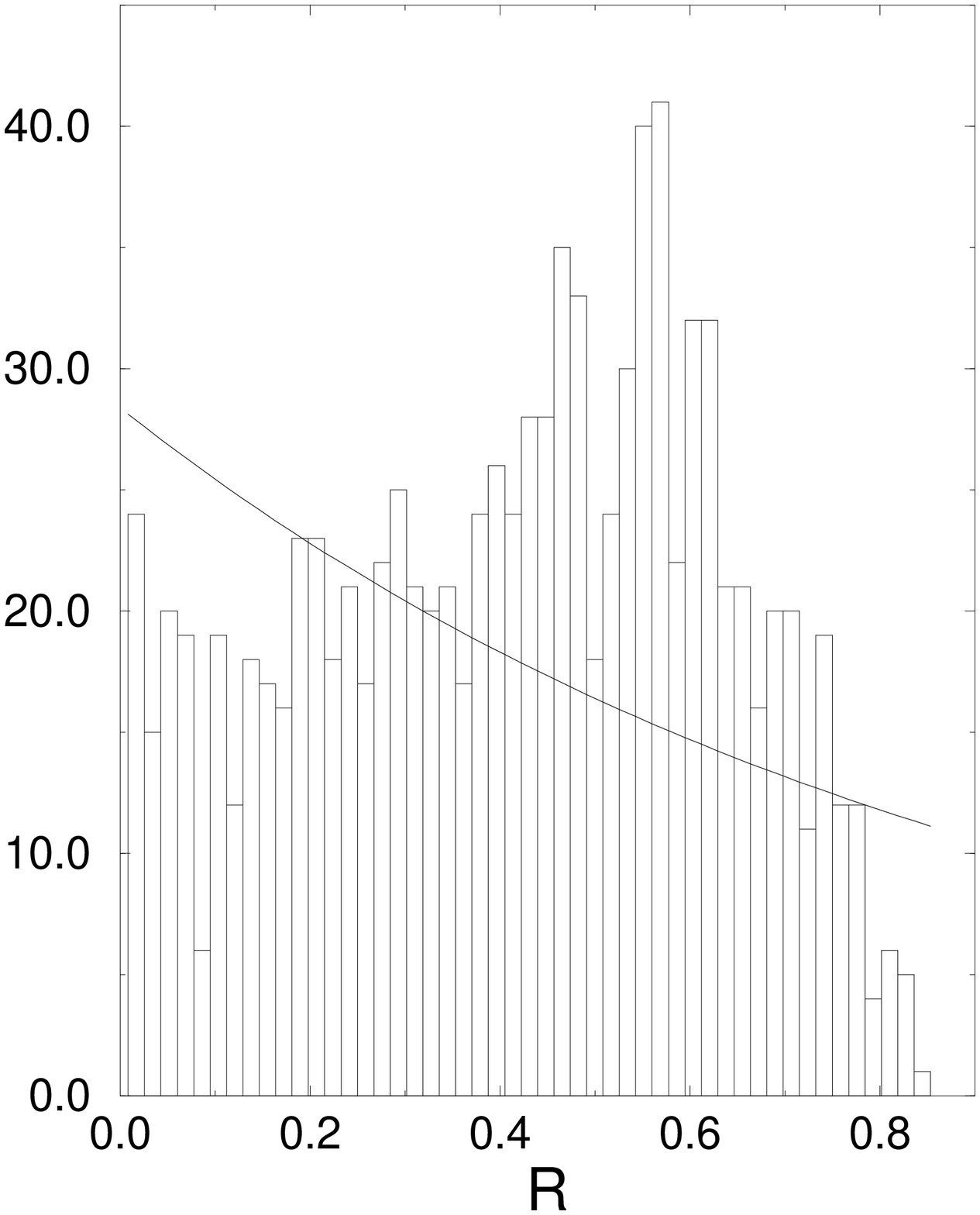}}
\caption{The distribution of the reflection coefficient $R$ 
for a cavity with 
one waveguide, supporting one channel: (left) in the presence of strong 
absorption ($\left\langle R \right\rangle =\alpha =0.034$); 
(right) for moderate absorption 
($\left\langle R \right\rangle =\alpha =0.410$).}
\label{pr50}
\end{figure}

Fig. 1 shows for $n=1$ the results of
the RMT numerical simulations (as histograms), compared with 
the present model for the corresponding value of $\alpha $ (continuous curves). 
For strong absorption the model works very well, the agreement with Rayleigh's 
law 
being excellent, while for moderate and weak absorptions the model fails.

Fig. 2 shows the distribution of the transmission coefficient $T$ 
obtained from a RMT simulation for $n=2$. The agreement with the  Rayleigh
distribution with the centroid $\left\langle T \right\rangle =\alpha /3$ is 
excellent;
$\left\langle R \right\rangle$ was also checked and found to agree with 
$2\alpha /3$.

That individual transmission and reflection coefficients attain a Rayleigh distribution 
for strong absorption can be understood as follows. $S_{ab}(E+iW)$ coincides 
\cite{mello-baranger99,mello-pereyra-seligman}
with the energy average of $S_{ab}(E)$ evaluated with a Lorentzian weighting function
of half-width $W$. If $\Gamma ^{corr}$ is the correlation energy, $W$ can be thought of
as containing $\sim m=W/\Gamma ^{corr}$ independent intervals.
If $m\gg 1$, by the central-limit theorem the real and imaginary parts of $S_{ab}$
attain a Gaussian distribution, and $\left|S_{ab}\right|^2$ an
exponential distribution. This seems to be the situation captured by the maximum-entropy 
approach. 

Summarizing, the results presented in this paper indicate that wave
scattering through classically chaotic cavities in the presence of strong
absorption can be described in terms of an information-theoretic model.
 
We benefited from the constructive criticism of the 
first version of the paper by C. W. J. Beenakker and P. W. Brouwer. 
One of the authors (PAM) acknowledges partial financial support from
CONACyT, Mexico, through Contract No. 2645P-E, as well as the hospitality of
Bar-Ilan University, where most of this work was performed. 

\begin{figure}[tbh]
\epsfxsize=4cm
\centerline{\epsffile{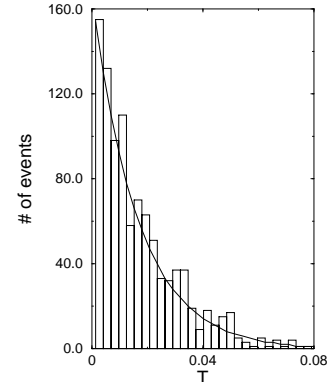}}
\caption{The distribution of the transmission coefficient 
$T$ 
for a cavity with two  waveguides, each supporting one open channel, in the 
presence of strong absorption ($\alpha =.049$).}
\label{nt}
\end{figure}

\end{multicols}

\end{document}